\documentclass{elsart}
\usepackage{natbib}
\begin{document}
\runauthor{Garrett, Morganti, Wrobel}
\begin{frontmatter}
\title{The Deepest \& Widest VLBI Survey yet: VLBA+GBT 1.4 GHz
  observations in Bootes.}
\author[mike]{Michael A. Garrett}
\author[joan]{Joan M. Wrobel}
\author[raff]{Raffaella Morganti}

\address[mike]{JIVE, Dwingeloo, NL}
\address[joan]{NRAO, Socorro, USA}
\address[raff]{ASTRON, Dwingeloo, NL}

\begin{abstract}
  
  We present preliminary results from the deepest VLBI observations yet
  conducted. VLBA+GBT 1.4~GHz observations of a region within NOAO-N,
  reach an r.m.s. noise level of 9 microJy per beam.  Three sources are
  clearly detected ($> 7\sigma$) within the inner 2 arcmins of the GBT
  primary beam, including two sub-mJy sources and the ``in-beam''
  calibrator. In addition, by tapering the data, we map out a much
  larger area of sky, reaching well beyond the half-power point of the
  GBT primary beam. An additional 6 sources are detected in the
  extended field. We comment briefly on the scientific motivation for
  even deeper and wider VLBI surveys, and note that the
  summed response of sources in the field will permit self-calibration
  techniques to be employed in {\it any} region of the radio sky,
  including so-called ``blank'' fields.

\end{abstract}

\end{frontmatter}

\section{Introduction}

Very Long Baseline Interferometry (VLBI) observations, and VLBI surveys
in particular, have made a significant contribution to our
understanding of radio galaxies and AGN quite generally. For example,
the discovery of superluminal motion in these systems provided the
first clue that orientation and viewing angle was an important
parameter in interpreting and eventually unifying different classes
of AGN, including radio galaxies and quasars (see Barthel \& van Bemmel
this volume). More recently, VLBI has provided direct evidence for an
evolutionary scenario in which very young compact radio sources are the
precursors of the giant extended radio sources associated with radio
galaxies.

VLBI centimeter radio source surveys are usually ``targeted'', that is
to say, that the sources observed are distributed randomly across the
sky and they are often pre-selected to be both bright ($S_{T} > 200$
mJy) and flat-spectrum ($\alpha < -0.5$) e.g. \cite{F2000}). Not surprisingly
such samples are largely dominated by moderate redshift ($z\sim 1-2$),
intrinsically luminous AGN.  In addition, the limited field-of-view
adopted by most observers, ensures that only one source is detected in
any given (snapshot) observation. VLBI surveying, is thus a slow
business and the biases introduced are substantial.

Over the past few years attempts have been made to survey much fainter
sources ($S_{T}>10$~mJy) using phase-reference techniques. Such surveys
take advantage of the huge catalogue of sources now available via the
FIRST VLA survey. These faint VLBI surveys are still targeted but they
are usually localised to one area of sky - a few square degrees that
includes a bright reference source surrounded by much fainter targets
\cite{G1999, W2002}. In this sense the surveys
are quite efficient (they minimize telescope slewing), and the
pre-selection criteria can be relaxed (often targets are only
pre-selected on their measured VLA size). The results of these surveys
are encouraging but one potential problem is that since the
observations are still targeted, the images are not very deep (the
total integration time per source is still only $\sim 15$ minutes).
Since the sources are not very bright (a peak flux of a few mJy is
typical) the dynamic range ($S_{peak}/S_{noise}$) is often $\sim 10$ or
less.  For brighter sources ($S_{peak} > 10$~mJy) the dynamic range is
often limited by errors introduced by conventional phase-referencing
(switching) techniques.  Under these circumstances its usually
difficult to classify the sources detected -- jets and other features
can easily be missed. In addition, it is quite clear that at these mJy
flux levels we are still probing, essentially the same radio source
population that the brighter surveys target too.

How can we make progress in this area ? The fundamental problem is that
the field-of-view of most VLBI images is quite limited, often
only a few hundred millarcseconds in extent. In this paper, we present
very preliminary results of the first deep, wide-field VLBI surveys
that attempt to expand the VLBI field-of-view by many  orders of
magnitude, so that it is only limited by the response of the primary
beam of the individual VLBI antennas. The idea is to demonstrate the
feasibility of imaging many dozens of target sources simultaneously
(just like short-baseline connected arrays) but with milliarcsecond
resolution and full sensitivity.

\begin{figure}[h]
\vspace{14cm}  
\includegraphics{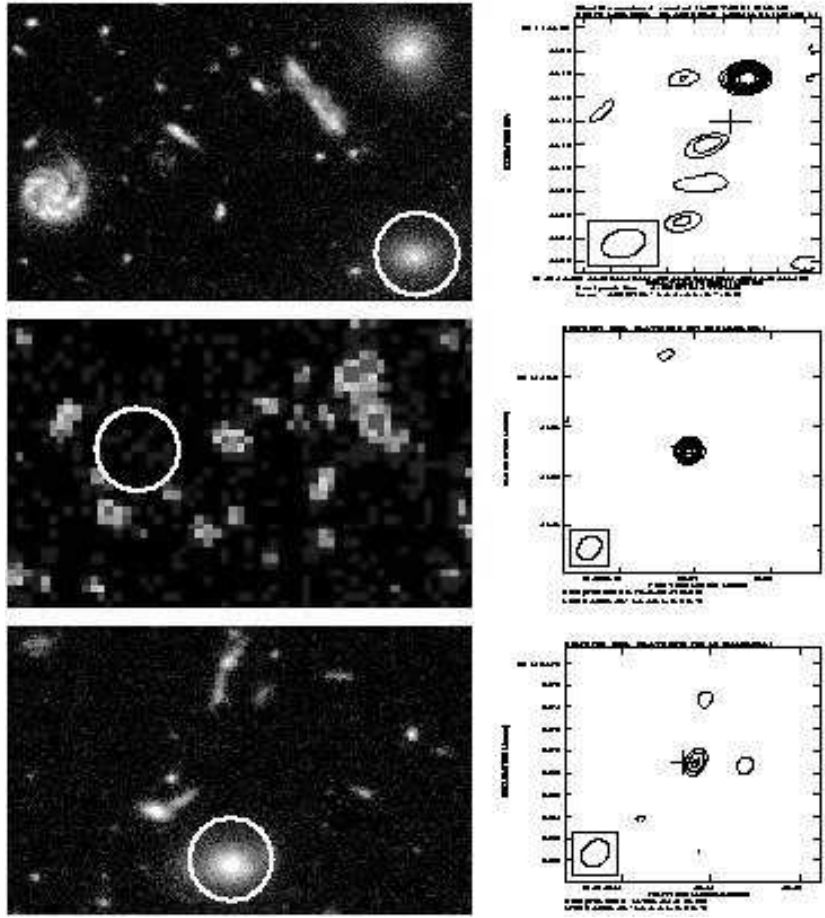}
\caption{EVN detections in the HDF: the distant
  z=1.01 FRI (top), the z=4.4 dusty obscured starburst hosting a hidden
  AGN (middle) and the faint 180 microJy, z=0.96 AGN (bottom). This
  latter source is the faintest radio source detected with the VLBI
  technique. Crosses represent the MERLIN-VLA positions for these
  sources.}
\label{evn_im}
\end{figure}

\section{Deep, Wide-field, VLBI Imaging} 

For a short baseline, connected element array, the field-of-view is
often set by the primary beam size of the individual telescope
elements. For VLBI this is hardly ever the case. In VLBI, a more
demanding limitation is set by the spectral resolution and time
sampling that are employed during data correlation. The resolution and
sampling must be fine enough to circumvent {\it both} bandwidth
smearing and time averaging effects, at later stages of the processing
(imaging) chain. In addition, since preserving the field-of-view scales
(computationally) with baseline length {\it squared}, the generation of
wide-field VLBI images is often limited by the off-line computing
resources available to the astronomer. This latter restriction, has
introduced another (psychological) barrier which is simply that most
VLBI practitioners are inclined to (over) average their data (both in
the time and frequency domains), in order to make even standard
continuum VLBI data sets more manageable. Data averaging collapses both
the field of view and thus the total information content, and leaves us
with the ``postage stamp'' VLBI images that we are all so familiar
with...

Recent attempts have been made to maintain the natural field-of-view
provided by VLBI correlators in order to image out much wider areas of
sky. Images a few arcminutes in extent can now be generated, and since
the full sensitivity of the array is brought to bear over a relatively
large field-of-view, several sources can be detected simultaneously.
Figure 1 shows the first, deep field VLBI observation of (what is
essentially) a blank (radio) field - the Hubble Deep
Field-North\cite{MAG2001}. In this paper, we report on a new attempts
to make deeper VLBI images, over a much wider area of sky.

\section{Deep, wide-field VLBA+GBT 1.4~GHz observations in Bootes
  (NOAO-N)}

The rms noise levels achieved by the EVN HDF-N observations were
limited by phase errors introduced via conventional, external
phase-referencing techniques. Some recent VLBA+GBT deep field
observations illustrate the gains to be made in employing ``in-beam''
phase referencing.  Figure 2 shows the deepest VLBI images made to date
(Garrett, Wrobel \& Morganti in prep).  The images (with an rms noise
of 9 microJy/beam in the centre of the field) were made from a 1.4~GHz
VLBA+GBT observing run ($3\times8$ hours, employing a sustained
recording data rate of 256~Mbps). In-beam phase-referencing was used to
provide essentially perfect phase corrections for this data set, and
eight sources are simultaneously detected ($> 7\sigma$) within and
outside the half-power point of the GBT primary beam. The response of
the ``in-beam'' calibrator was subtracted from the full data set.  Of
the eight sources detected, two sub-mJy sources are located within the
primary beam of the GBT, in addition to the in-beam phase reference
calibrator (a compact 20 mJy source \cite{W2002}). The images of
sources far from the field centre are tapered -- the temporal and
spectral resolution of the data is only adequate for sources that lie
within the inner 2 arcmins of the primary beam of the GBT.

\begin{figure}[t]
\vspace{13cm}  
\includegraphics{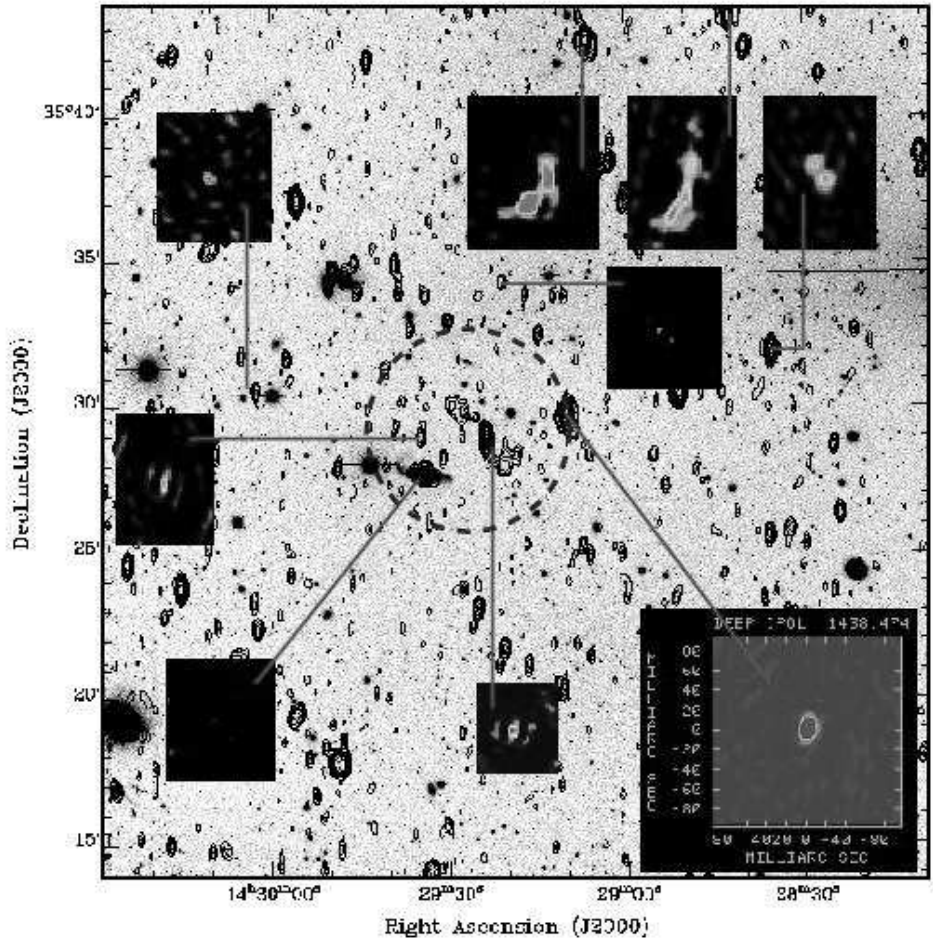}
\caption{Deep VLBA+GBT 1.4 GHz observations of a small portion of the
  NOAO-N Bootes deep field. The VLBI detections are shown inset. Radio
  line contours (produced by the WSRT) are superimposed on the NOAO
  optical field). One non-detection is also shown (bottom left) - a
  bright, presumably nearby (star forming) spiral galaxy that is well
  detected by the WSRT. These are the deepest images made with VLBI to
  date (Garrett, Wrobel \& Morganti in preparation) with an r.m.s.
  noise of $9\mu$Jy/beam.}
\end{figure}

The total (target) data set size is 60 Gbytes (0.5 secs integration,
$1024 \times 62.5$~kHz channels). Images were made with the AIPS task
{\sc IMAGR} - dirty maps/beams of each sub-band (IF) for each of the
three epochs were generated blindly, and then simply co-added together.
Many ``small'' patches ($6^{\prime\prime}\times 6^{\prime\prime}$)
of the field were imaged, based on the positions of a deep, complimentary
WSRT 1.4 GHz survey \cite{RMMAG02}. The latter survey uses the upgraded
WSRT system and reaches an r.m.s. noise level of $\sim 13$~$\mu$Jy/beam
 in a 12 hour observing run. The computational task of generating a
map of each patch of sky is considerable -- about 8 hours was required
to produce each dirty image (a dual processor, 2 GHz, Linux
box was employed). The analysis of these data is on-going. 
For sources that were bright enough, CLEAN maps were produced by simply
subtracting the dirty beam from the dirty image (AIPS task APCLN). More
complicated tasks (e.g. IMAGR) involving a visibility based CLEAN are
currently prohibitively expensive in terms of CPU requirements.

\vspace{-0.5cm} 
\section{The Nature of the high-z, obscured sub-mm/radio source
  population and Future technical advances in VLBI} 

At full resolution and maximum sensitivity, our 1.4~GHz VLBA+GBT
observations, can detect radio sources with a brightness temperature in
excess of $5 \times 10^{5}$K. The sources we detect must therefore be
powered by AGN activity, rather than extended star formation processes
\cite{Cond92}. For example, we do not detect compact emission from a
nearby spiral galaxy (NGC 5646, $z \sim 0.03$), although it is one of
the brightest ($S_{T}\sim 3$~mJy) sources in the GBT primary beam (see
bottom left hand corner of Figure 2). The fact that the WSRT radio
emission follows the optical isophotes, and that this source obeys the
FIR/radio correlation, strongly suggests the radio emission arises
mainly from star formation. All VLBI detections presented in Figure 2,
represent {\it direct} evidence for AGN activity.

The long term motivation for these deep VLBI surveys is to determine a
lower limit to the contribution AGN make to the faint radio source
population in general, and the optically faint (obscured) radio and
sub-mm (SCUBA) source population in particular. In order to make any
impact in this area, it is essential that many sources are surveyed
over large areas of sky, to ($1\sigma$) depths of a few microJy. In
principle, global VLBI arrays, employing disk-based
recording\cite{MAG2003} can reach these kind of sensitivity levels. A
programme to harness the full capacity of the EVN correlator at JIVE
(PCInt) will lead this year, to a remarkable expansion in the
field-of-view accessible to VLBI observers. Output data rates of up to
160 MBytes/sec, will permit milliarcsecond imaging of huge swathes of
sky, limited only by the primary beam of individual VLBI telescope
elements.  Since it will be possible to simultaneously sample the {\it
  summed} response of all compact radio sources within (and indeed
beyond) the half-power point of the VLBI telescope primary beam, simple
self-calibration of the target field will {\it always} be possible!
Access to GRID like computing resources may be the best way to analyse
the huge data sets generated.  PCInt will permit dozens of sources to be
detected simultaneously, and imaged at milliarcsecond resolution with
full sensitivity. In this way, huge, unbiased VLBI surveys will be
conducted, the bulk of the targets being faint sources with flux
densities of only a few tens of microJy. Some of these will include the
distant, high-z population of dust obscured systems.

\end{document}